\documentclass[a4j]{article}
\usepackage[top=25truemm,bottom=25truemm,left=25truemm,right=25truemm]{geometry}
\usepackage[english]{babel}
\usepackage{authblk}
\usepackage{subfigure}
\usepackage{wrapfig}
\usepackage{graphicx}

\begin{document}

\title{Optimization of layer composition for ILD ECAL}
\author[1]{Hiroki Sumida}
\author[1]{Masato Matama}
\author[1]{Yuji Sudo}
\author[1]{Taikan Suehara}
\author[1]{Tamaki Yoshioka}
\author[1]{Kiyotomo Kawagoe}
\author[2]{Daniel Jeans}
\author[3]{Tohru Takeshita}
\author[3]{Katsushige Kotera}

\affil[1]{Faculty of Science, Kyushu University}
\affil[2]{International Center for Elementary Particle Physics, The University of Tokyo}
\affil[3]{Faculty of Science, Shinshu University}

\date{}
\maketitle

\begin{abstract}
\noindent
International Large Detector (ILD) adopts Particle Flow Algorithm (PFA) for precise measurement of multiple jets. The electromagnetic calorimeter (ECAL) of ILD has two candidates sensor technologies for PFA, which are pixelized silicon sensors  and scintillator-strips with silicon photomultipliers. Pixelized silicon sensors have higher granularity for PFA, however they have  an issue of cost reduction. In contrast, scintillator-strips have an advantage of relatively low cost and a disadvantage of degradation of position resolution by ghost hits, which are generated by orthogonal arrangement. Hybrid ECAL using both candidates is proposed to supplement these disadvantages. In this paper, we report an optimization study of the hybrid ECAL using detector simulation.
\end{abstract}

\footnote[0]{Talk presented at the International Workshop on Future Linear Colliders (LCWS15), Whistler, Canada, 2-6 November 2015.}

\section{Introduction}

International Linear Collider (ILC) is a future lepton collider which is expected to be constructed in Japan [1]. It can measure higgs boson and top quarks precisely. This measurement is useful to search for new particles and physics. International Large Detector (ILD) [2], which is one of the detector concepts in ILC, have to have suitable structures for Particle Flow Algorithm (PFA) because ILD needs to measure multiple jets contained in final states of these physics processes of ILC precisely. 

PFA is a method to identify particles in a jet individually and to measure momentum and energy at optimal detectors for each particle. Jet energy is measured by mainly trackers, ECAL and hadron calorimeter (HCAL). At the trackers, momentum of  charged particles is measured, where energies of photons and neutral hadrons are measured at ECAL and HCAL, respectively. 

ILD ECAL has two candidates of sensitive detectors (Figure 1). One is silicon semiconductor sensor with $5.5\times5.5$ mm$^2$ pixels which has high granularity for PFA, however its cost reduction is an important issue. The other is scintillator-strips with silicon photomultipliers as photon detectors. Its cost is relatively low, however jet energy resolution is worse by ghost hits which are made by orthogonal arrangement of $45\times5$ mm$^2$ strip when there are more than two hits on $45\times45$ mm$^2$ area spontaneously.

We optimized a hybrid ECAL concept, which uses both silicon and scintillator detectors to reduce ECAL cost and maintain its performance.

\begin{figure}[htbp]
\begin{minipage}{0.5\hsize}
\centering
\includegraphics[width=2.8in, bb= 0 0 3264 2448]{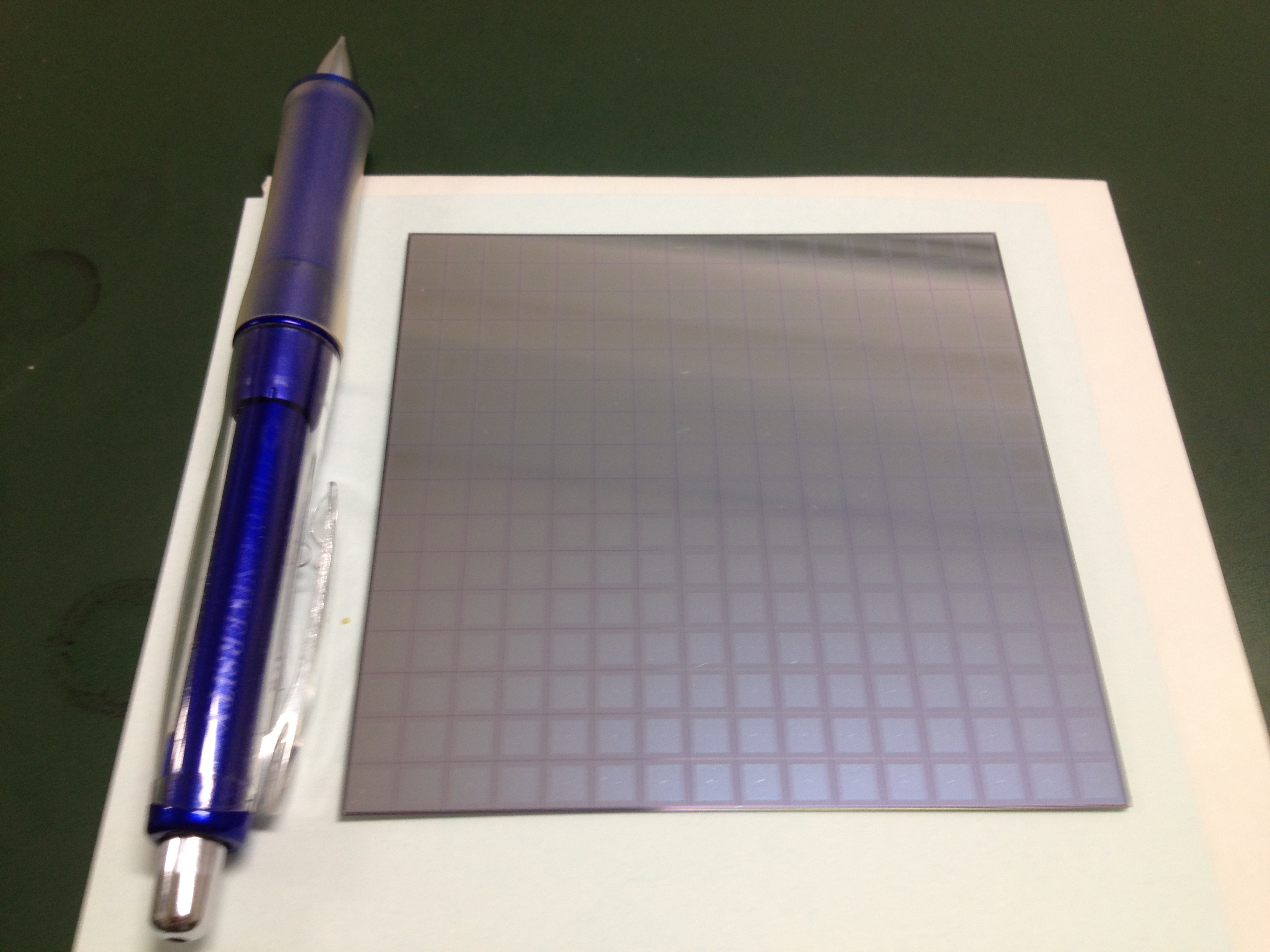}
\end{minipage}
\begin{minipage}{0.5\hsize}
\centering
\includegraphics[width=2.8in, bb= 0 0 1024 703]{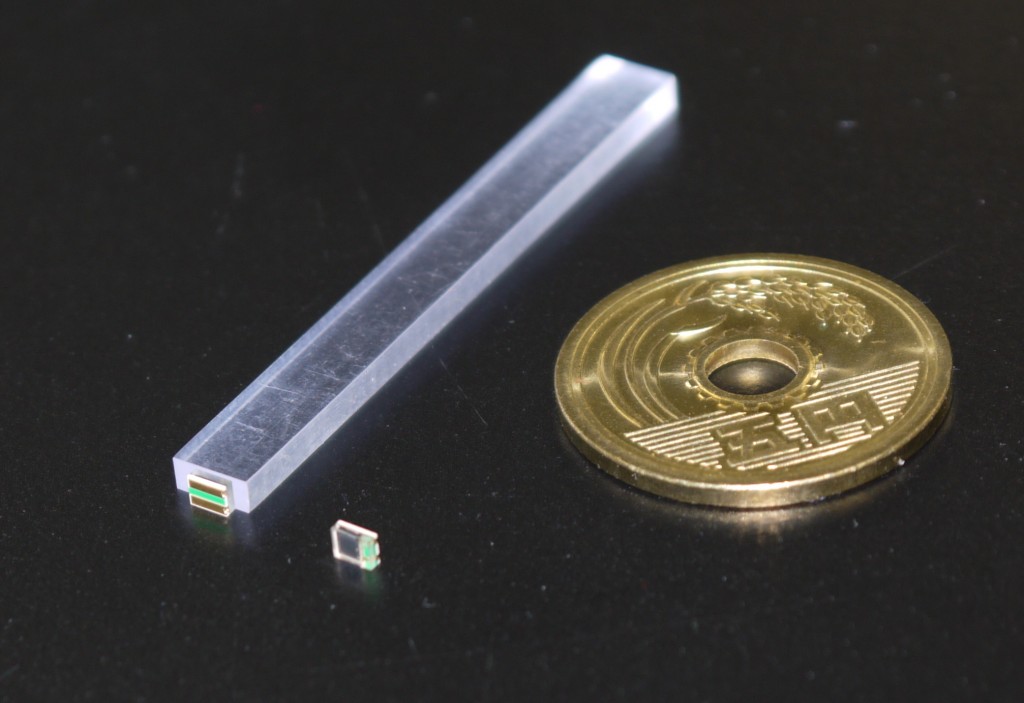}
\end{minipage}
\caption{Sensitive detectors of ILD ECAL. Left is pixelized silicon semiconductor and right is scintillator-strip with MPPC.}
\end{figure}

\newpage
\section{Optimization of hybrid ECAL}

Recent baseline design of the ILD ECAL is described in the Detailed Baseline Design (DBD) report. The structure uses pixelized silicon semiconductor as sensitive detectors which can reach high energy resolution. However, it causes high occupation of cost ratio in the whole ILD cost. To deal with this problem, two ideas are proposed. One is to reduce the whole size of ILD, and the other is to use hybrid ECAL instead of ECAL with all silicon semiconductor. 

This paper describes optimization study of hybrid structure with detector simulation.

\subsection{Simulation framework}

ILCSoft v01\_16\_02 is a software suite including numerous packages for physics analysis and detector optimization of ILC, used for the following analysis. For the event generation, Whizard and Pythia packages are used for di-jet events. Single particles are generated within the detector simulation software with particle guns. Mokka is a detector simulation package based on GEANT4. ILD\_o1\_v5 geometry in the Mokka framework is used for the detector simulation. After the simulation, we performed the ILD standard event reconstruction, including digitization of the hits, tracking and track fitting, and PandoraPFA for particle flow.

\subsection{Setup of simulation}

This study focuses on the hybrid ECAL with reasonable cost and performance. As an example, we set the silicon sensors in first 14 detector layers (inner region) and scintillator-tiles in the rest of layers (outer region) (Figure 2). The number of silicon layers is fixed because we would like to restrict the cost of silicon. For the scintillator, we adopt $15\times15$ mm$^2$ tiles. Comparison between tiles and strips is not in the scope of this study and should be studied separately.

\newpage
\begin{figure}[htbp]
\centering
\includegraphics[width=5in, bb= 0 0 966 493]{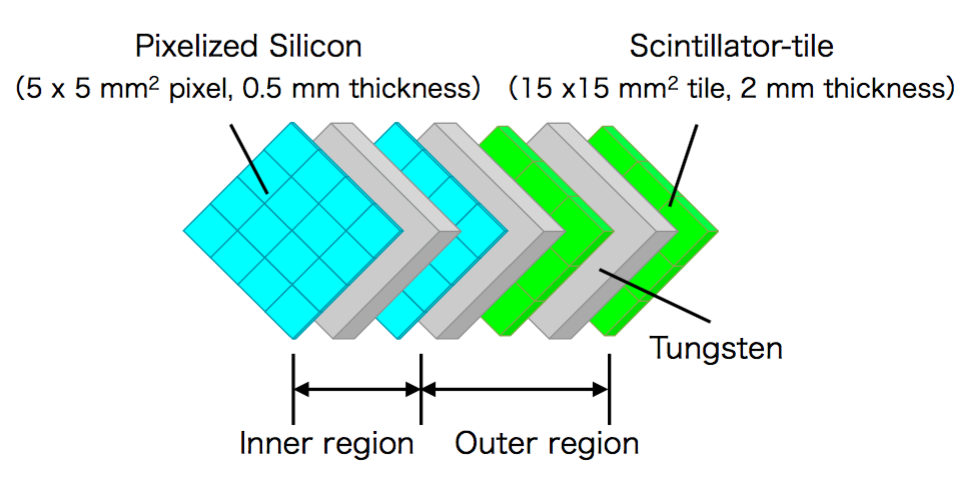}
\caption{Configuration of hybrid ECAL}
\end{figure}

In this study, we change the boundary between inner and outer region and the number of scintillator layers, but whole thickness of absorber layers is fixed ($22.8X_0=79.8$ mm). Considered configurations are listed in Figure 3 and Table 1-4.

As a performance comparison, we investigated energy resolution of di-jet ($e^+e^-\to Z\to q\bar{q}$)  events of various center-of-mass energies. We use RMS90 energy resolution obtained by PandoraPFA [3], and PerfectPFA which utilizes MC information for clustering instead of real particle flow. Confusion term is defined as squared difference of PandoraPFA and PerfectPFA.

\begin{figure}[htbp]
\centering
\includegraphics[width=3in, bb= 0 0 694 533]{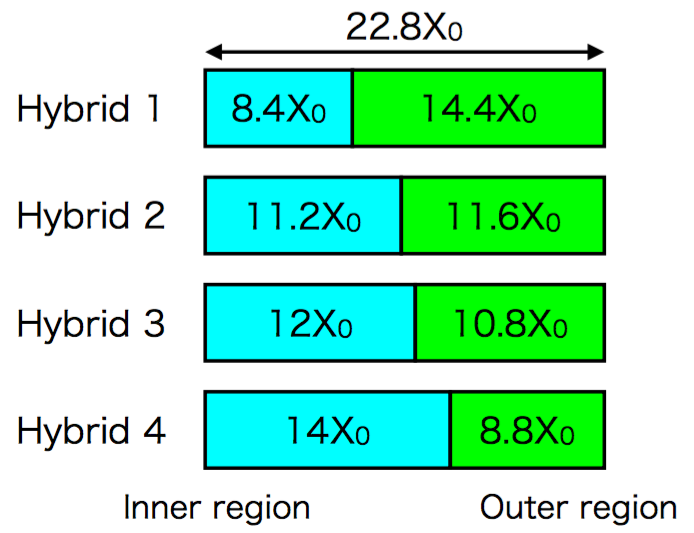}
\caption{Location of a boundary between inner and outer region}
\end{figure}

\begin{table}[htbp]
\begin{center}
\begin{tabular}{|c|c|c|c|c|}
\hline
Configuration & Boundary & Si & Sc & Absorber (Inner/Outer) \\ \hline\hline
Hybrid (1-1) & $8.4X_0$ & 14 layers & 20 layers & 2.1 mm$\times$14 layers/2.653 mm$\times$19 layers\\
Hybrid (1-2) & $8.4X_0$ & 14 layers & 22 layers & 2.1 mm$\times$14 layers/2.4 mm$\times$21 layers\\
Hybrid (1-3) & $8.4X_0$ & 14 layers & 24 layers & 2.1 mm$\times$14 layers/2.191 mm$\times$23 layers\\
Hybrid (1-4) & $8.4X_0$ & 14 layers & 26 layers & 2.1 mm$\times$14 layers/2.016 mm$\times$25 layers\\ \hline
\end{tabular}
\caption{Configuration of ECAL at a boundary of $8.4X_0$}
\end{center}
\end{table}

\begin{table}[htbp]
\begin{center}
\begin{tabular}{|c|c|c|c|c|}
\hline
Configuration & Boundary & Si & Sc & Absorber (Inner/Outer) \\ \hline\hline
Hybrid (2-1) & $11.2X_0$ & 14 layers & 16 layers & 2.8 mm$\times$14 layers/2.707 mm$\times$15 layers\\
Hybrid (2-2) & $11.2X_0$ & 14 layers & 18 layers & 2.8 mm$\times$14 layers/2.388 mm$\times$17 layers\\
Hybrid (2-3) & $11.2X_0$ & 14 layers & 20 layers & 2.8 mm$\times$14 layers/2.137 mm$\times$19 layers\\
Hybrid (2-4) & $11.2X_0$ & 14 layers & 22 layers & 2.8 mm$\times$14 layers/1.933 mm$\times$21 layers\\ \hline
\end{tabular}
\caption{Configuration of ECAL at a boundary of $11.2X_0$}
\end{center}
\end{table}

\begin{table}[htbp]
\begin{center}
\begin{tabular}{|c|c|c|c|c|}
\hline
Configuration & Boundary & Si & Sc & Absorber (Inner/Outer) \\ \hline\hline
Hybrid (3-1) & $12X_0$ & 14 layers & 16 layers & 3.0 mm$\times$14 layers/2.52 mm$\times$15 layers\\
Hybrid (3-2) & $12X_0$ & 14 layers & 18 layers & 3.0 mm$\times$14 layers/2.224 mm$\times$17 layers\\
Hybrid (3-3) & $12X_0$ & 14 layers & 20 layers & 3.0 mm$\times$14 layers/1.989 mm$\times$19 layers\\
Hybrid (3-4) & $12X_0$ & 14 layers & 22 layers & 3.0 mm$\times$14 layers/1.8 mm$\times$21 layers\\ \hline
\end{tabular}
\caption{Configuration of ECAL at a boundary of $12X_0$}
\end{center}
\end{table}

\begin{table}[htbp]
\begin{center}
\begin{tabular}{|c|c|c|c|c|}
\hline
Configuration & Boundary & Si & Sc & Absorber (Inner/Outer) \\ \hline\hline
Hybrid (4-1) & $14X_0$ & 14 layers & 16 layers & 3.5 mm$\times$14 layers/2.053 mm$\times$15 layers\\
Hybrid (4-2) & $14X_0$ & 14 layers & 18 layers & 3.5 mm$\times$14 layers/1.812 mm$\times$17 layers\\
Hybrid (4-3) & $14X_0$ & 14 layers & 20 layers & 3.5 mm$\times$14 layers/1.621 mm$\times$19 layers\\
Hybrid (4-4) & $14X_0$ & 14 layers & 22 layers & 3.5 mm$\times$14 layers/1.467 mm$\times$21 layers\\ \hline
\end{tabular}
\caption{Configuration of ECAL at a boundary of $14X_0$}
\end{center}
\end{table}

\newpage
\subsection{Optimization of number of scintillator layers}

Figure 4-7 shows the dependence on number of scintillator layers in hybrid 1-4 geometries. These plots show that there are no significant differences among the variation of the number of scintillator layers contrary to an expectation that increasing of the number of scintillator layers makes the resolution better. This may caused by calibration procedure or different thicknesses between silicon and scintillator layers. This will be investigated in future studies.

\begin{figure}[htbp]
\begin{minipage}{0.5\hsize}
\centering
\includegraphics[width=2.5in, bb = 0 0 545 567]{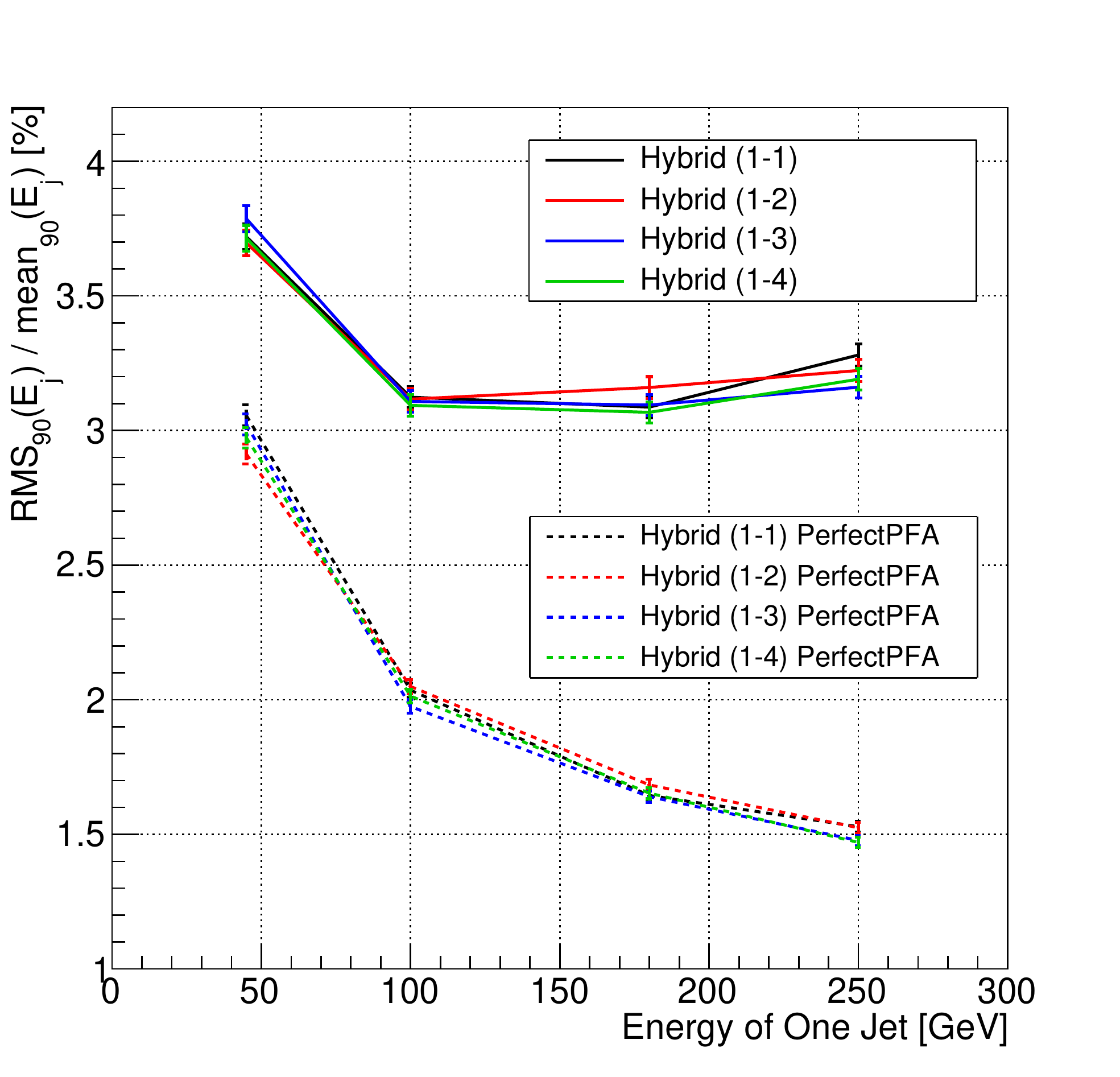}
\end{minipage}
\begin{minipage}{0.5\hsize}
\centering
\includegraphics[width=2.5in, bb = 0 0 545 567]{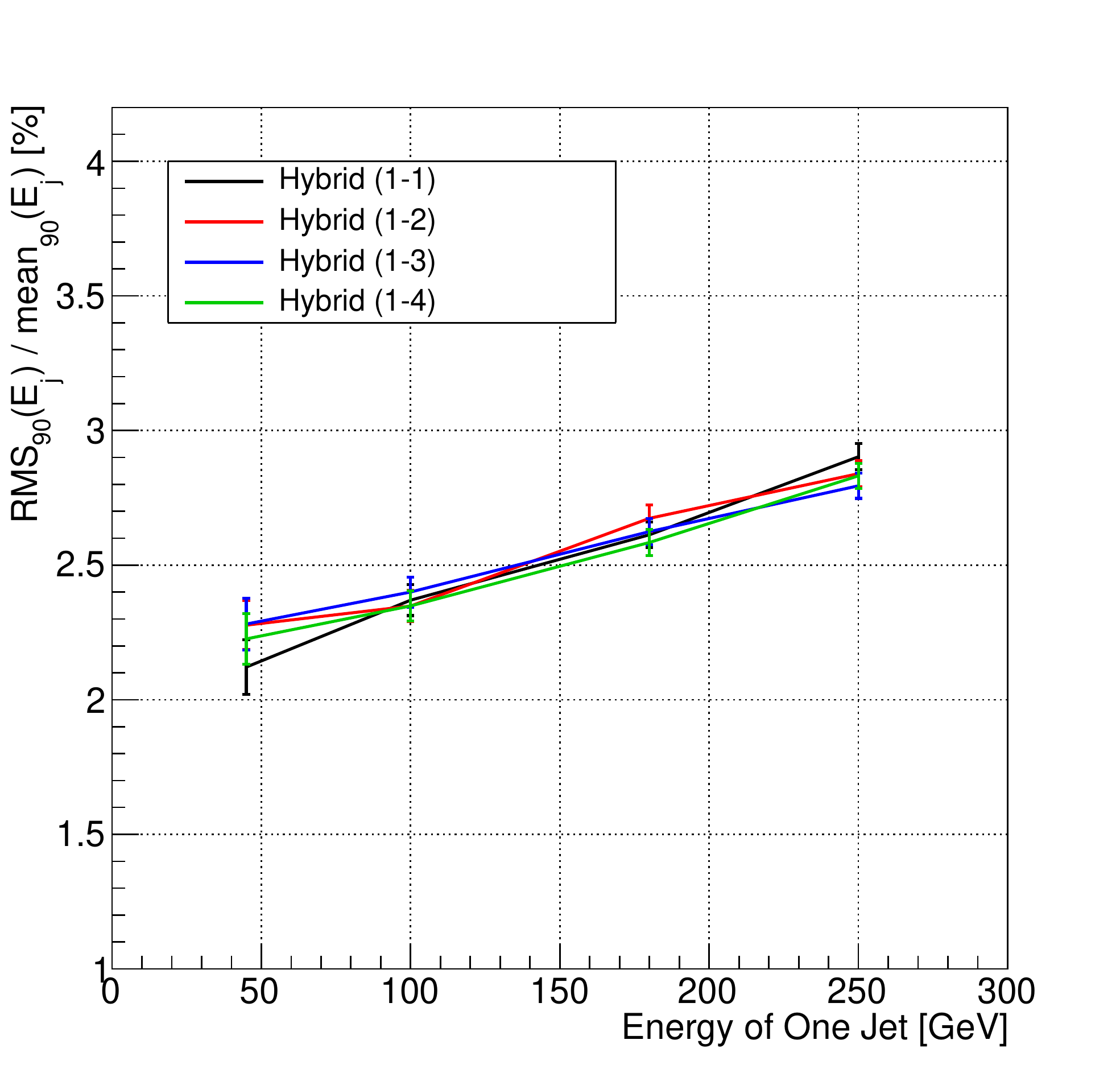}
\end{minipage}
\caption{Energy resolution (left) and confusion term (right) of configurations with a boundary of $8.4X_0$}
\end{figure}

\begin{figure}[htbp]
\begin{minipage}{0.5\hsize}
\centering
\includegraphics[width=2.5in, bb = 0 0 545 567]{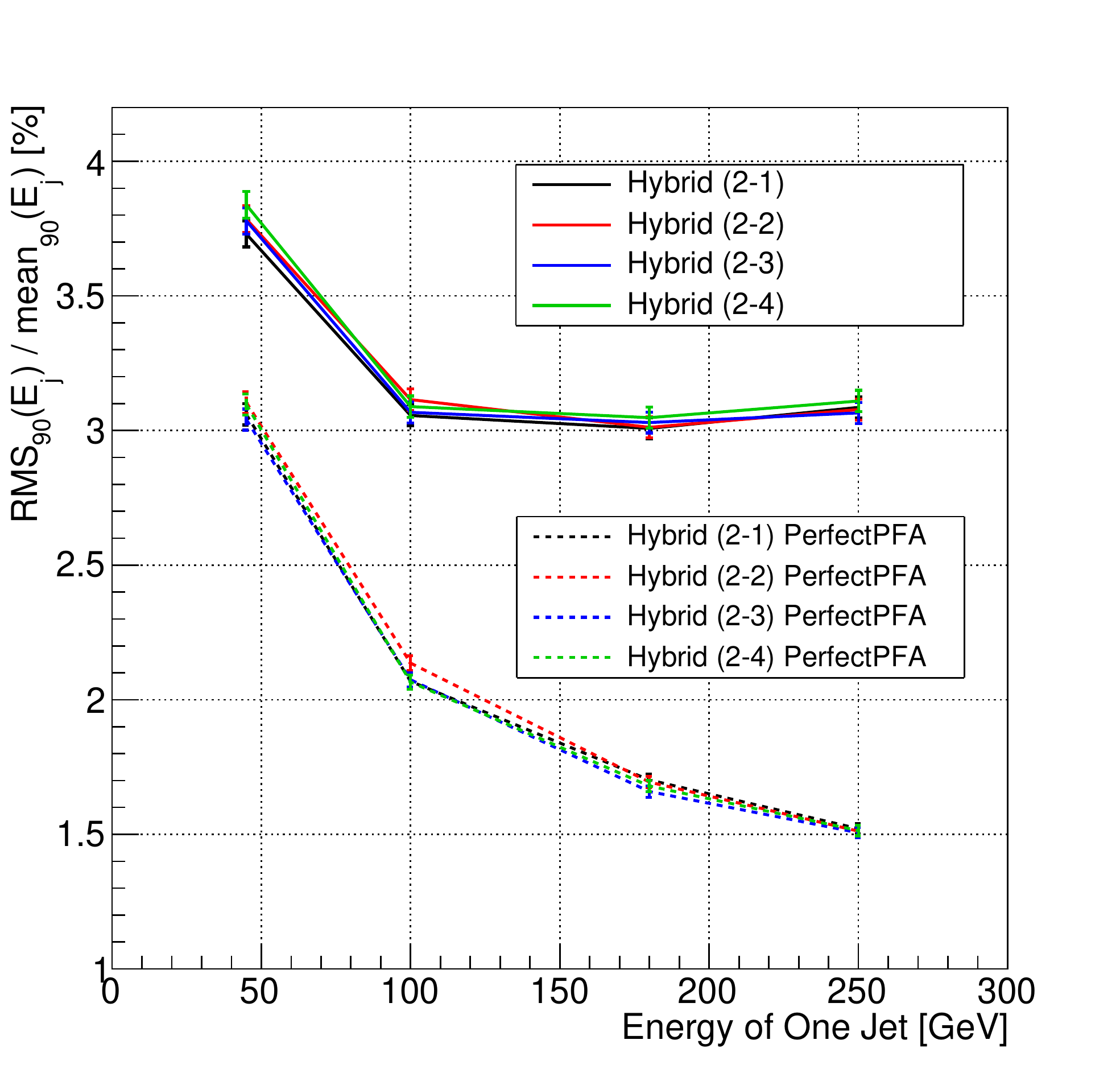}
\end{minipage}
\begin{minipage}{0.5\hsize}
\centering
\includegraphics[width=2.5in, bb = 0 0 545 567]{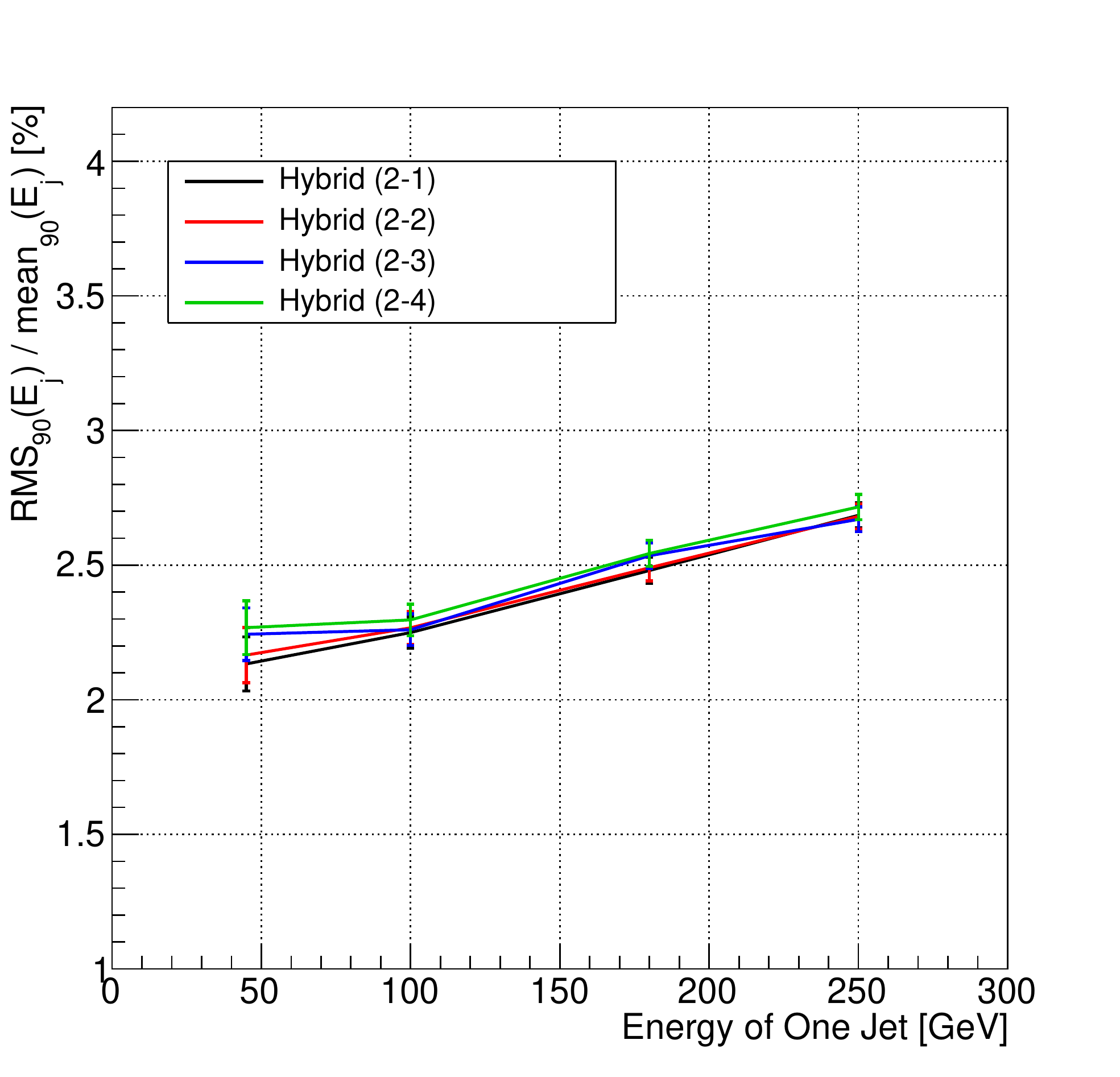}
\end{minipage}
\caption{Energy resolution (left) and confusion term (right) of configurations with a boundary of $11.2X_0$}
\end{figure}

\begin{figure}[htbp]
\begin{minipage}{0.5\hsize}
\centering
\includegraphics[width=2.5in, bb = 0 0 545 567]{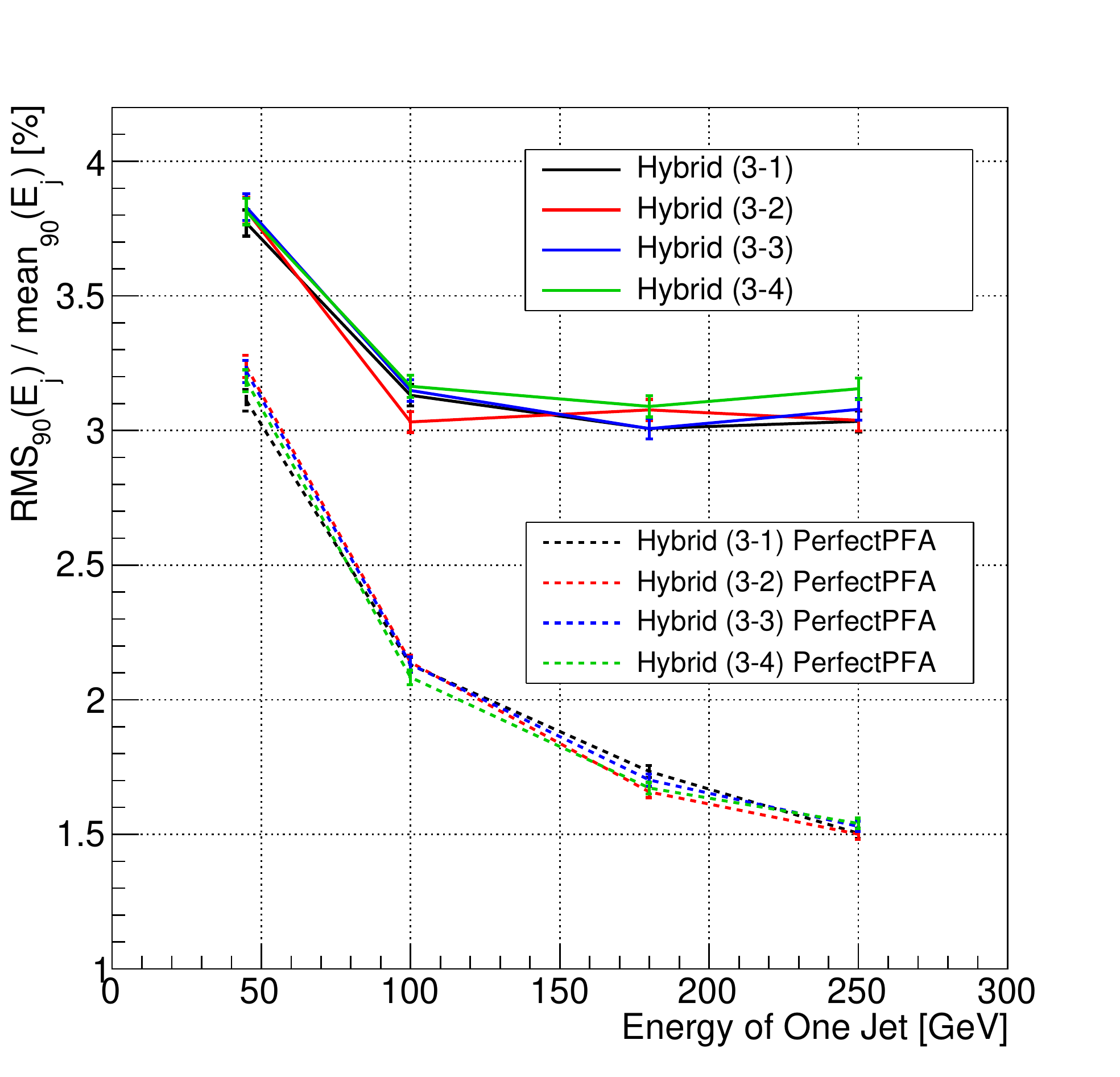}
\end{minipage}
\begin{minipage}{0.5\hsize}
\centering
\includegraphics[width=2.5in, bb = 0 0 545 567]{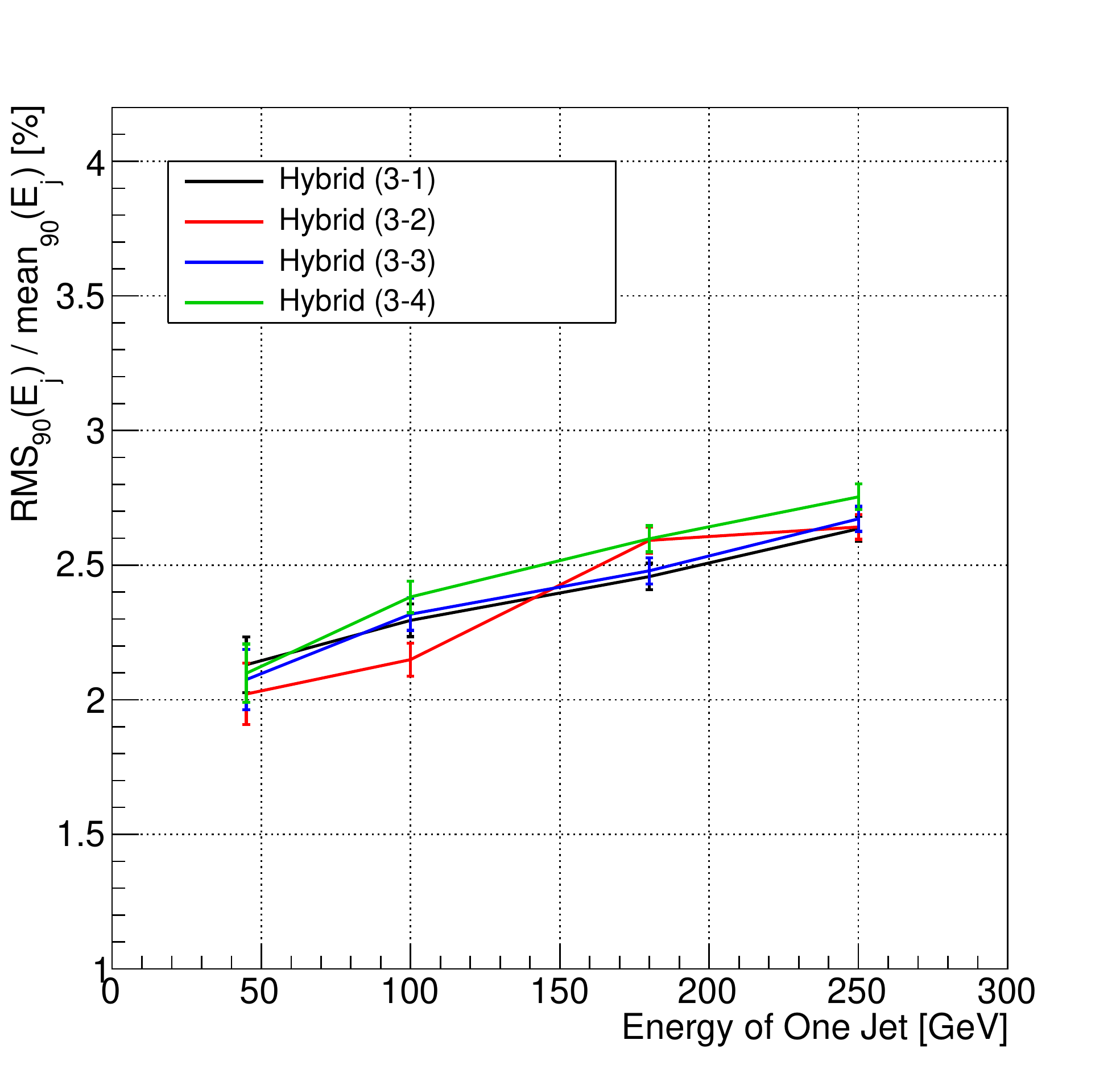}
\end{minipage}
\caption{Energy resolution (left) and confusion term (right) of configurations with a boundary of $12X_0$}
\end{figure}

\begin{figure}[htbp]
\begin{minipage}{0.5\hsize}
\centering
\includegraphics[width=2.5in, bb = 0 0 545 567]{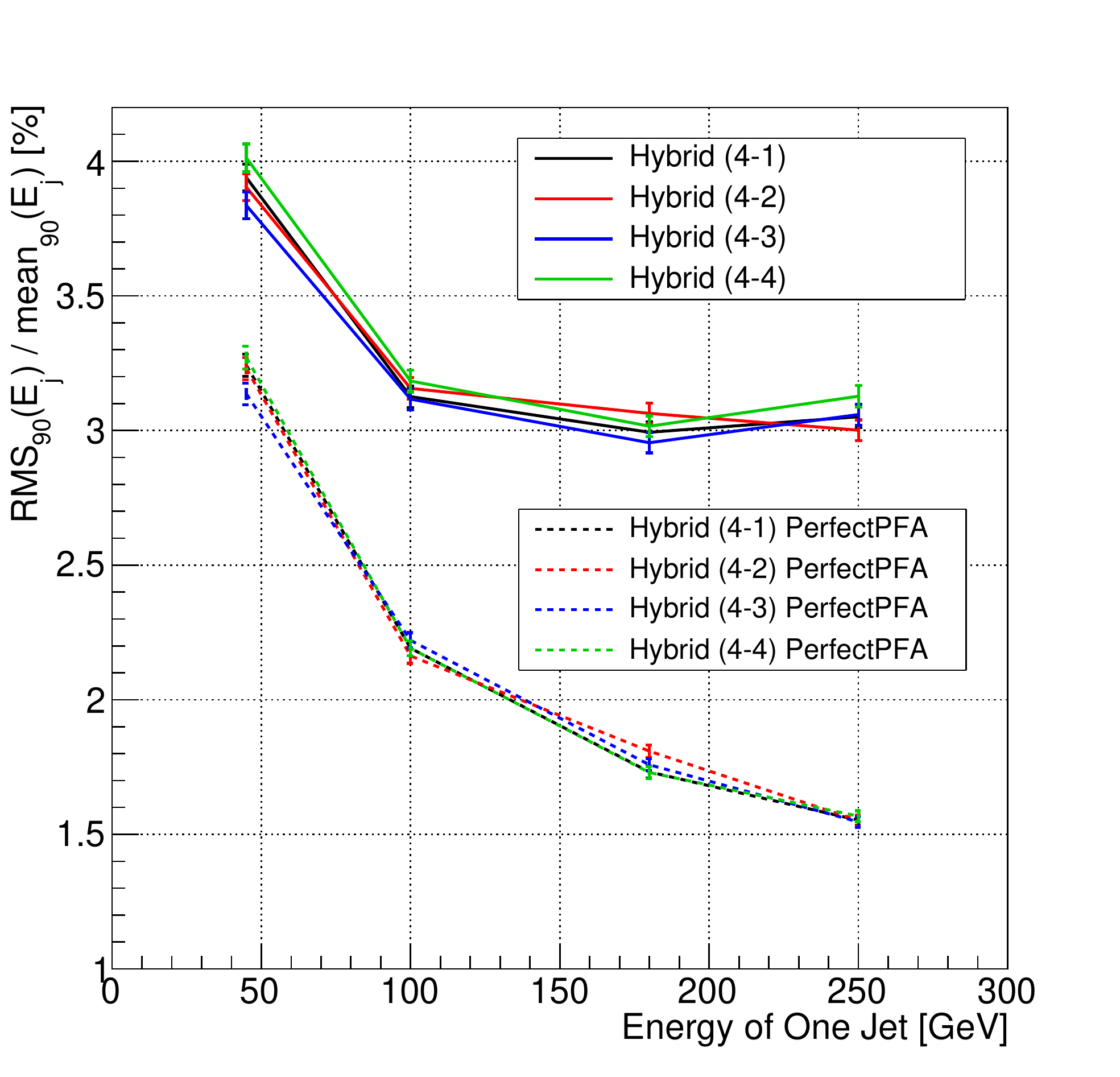}
\end{minipage}
\begin{minipage}{0.5\hsize}
\centering
\includegraphics[width=2.5in, bb = 0 0 545 567]{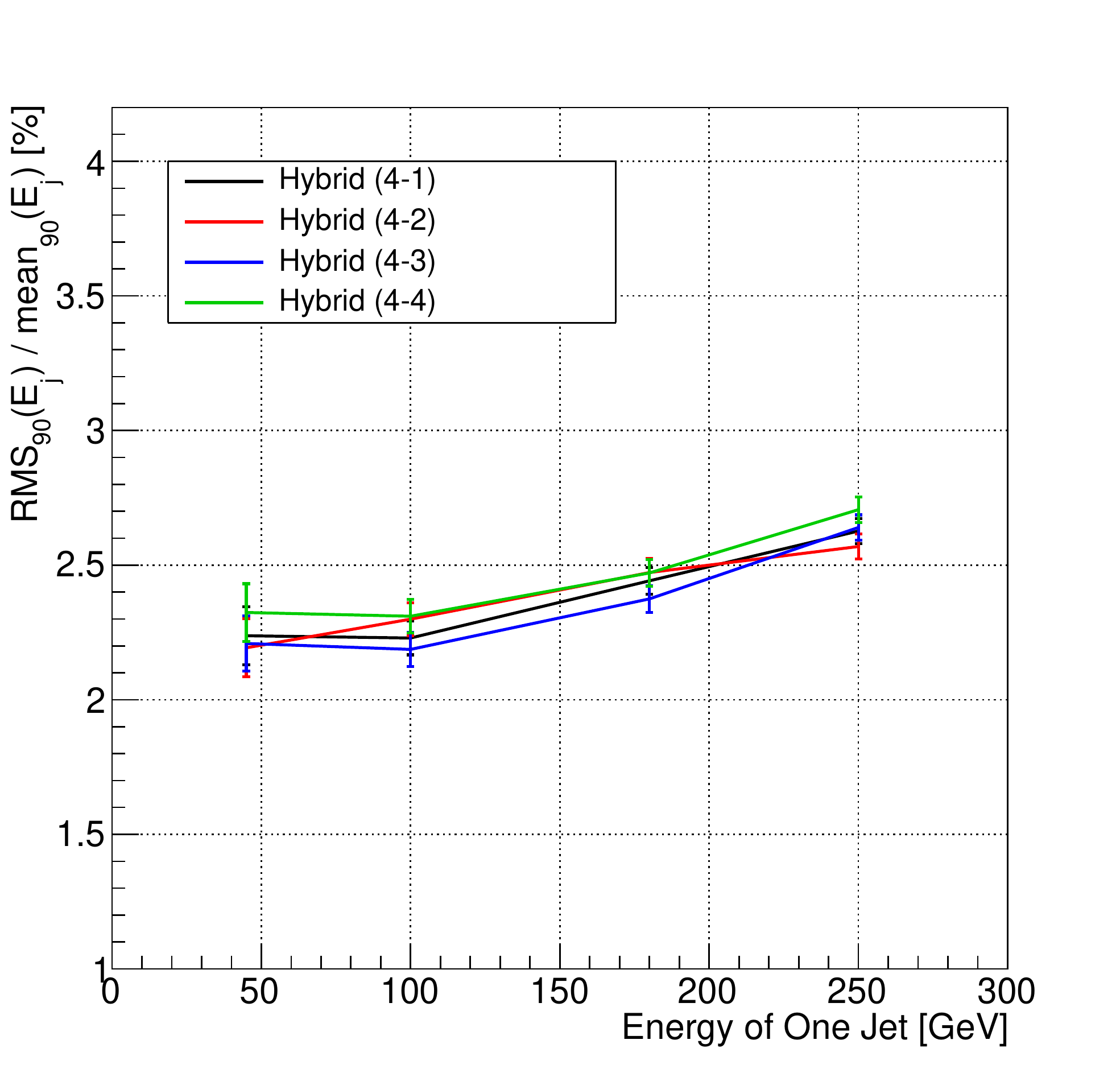}
\end{minipage}
\caption{Energy resolution (left) and confusion term (right) of configurations with a boundary of $14X_0$}
\end{figure}

\newpage
\subsection{Comparison with DBD configuration}

We compared the jet energy resolution of considered configurations with that of silicon ECAL in DBD. For this comparison, we picked up 4 considered configurations which had the least number of scintillator layers in each boundary (Hybrid(1-1), Hybrid(2-1), Hybrid(3-1), Hybrid(4-1)) because these configurations has a possibility of the biggest cost reduction. Figure 8 shows the result of jet energy resolution and confusion term. It is seen that energy resolutions of PandoraPFA with the boundary in outer direction (Hybrid(2-1), (3-1), (4-1)) are almost same as that of DBD structure at high energy. The reason of degradation of hybrid configurations with respect to DBD configuration is not identified and should be investigated further. The configuration with the boundary of $11.2X_0$ (Hybrid(2-1)) has average degradation of energy resolution of 2.7\% with reducing the cost of about 30\% compared with silicon ECAL in DBD. For the confusion term, DBD configuration has slightly better performance reflecting the difference in the pixel size.

\begin{figure}[htbp]
\begin{minipage}{0.5\hsize}
\centering
\includegraphics[width=2.8in, bb = 0 0 545 567]{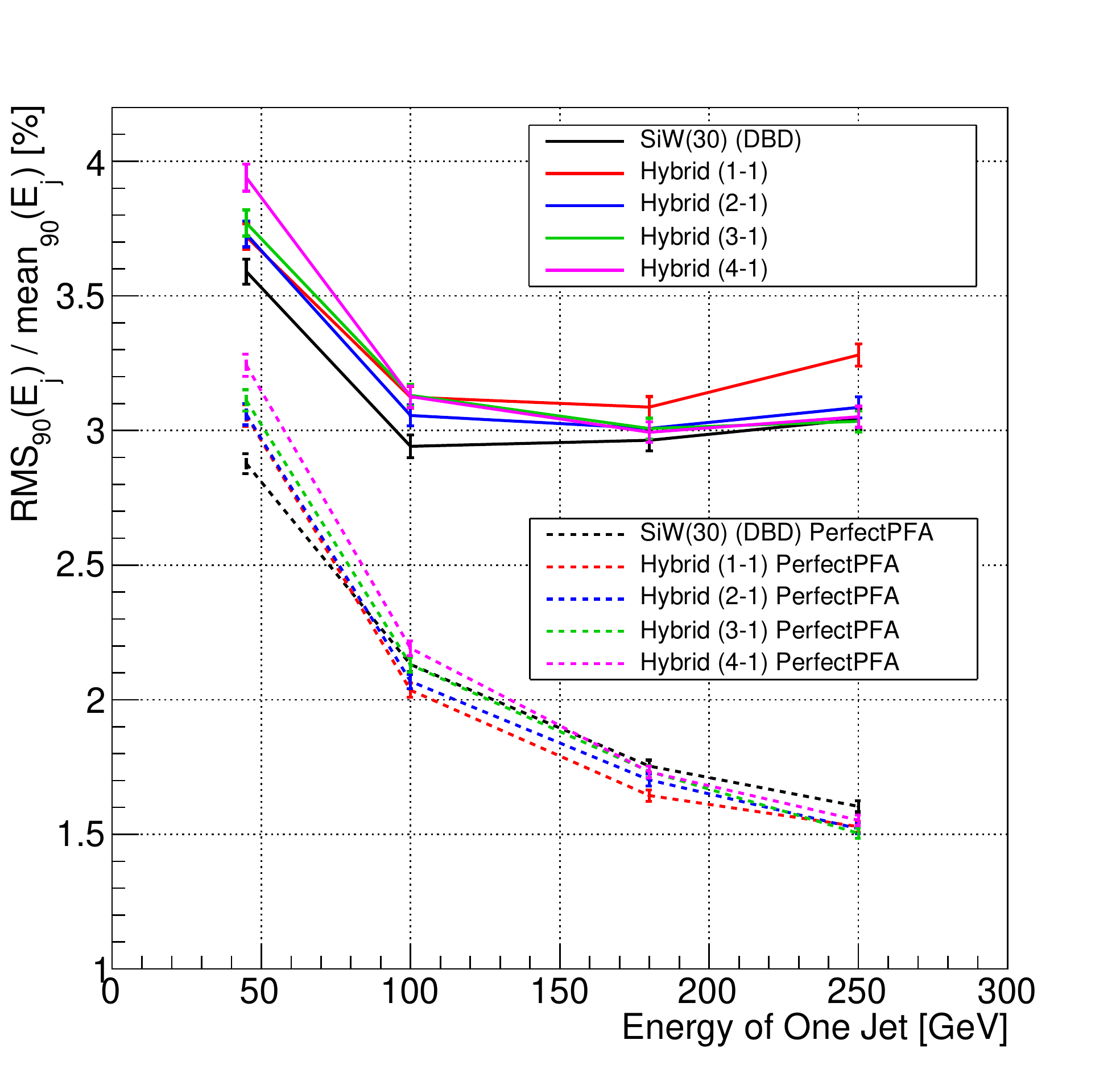}
\end{minipage}
\begin{minipage}{0.5\hsize}
\centering
\includegraphics[width=2.8in, bb = 0 0 545 567]{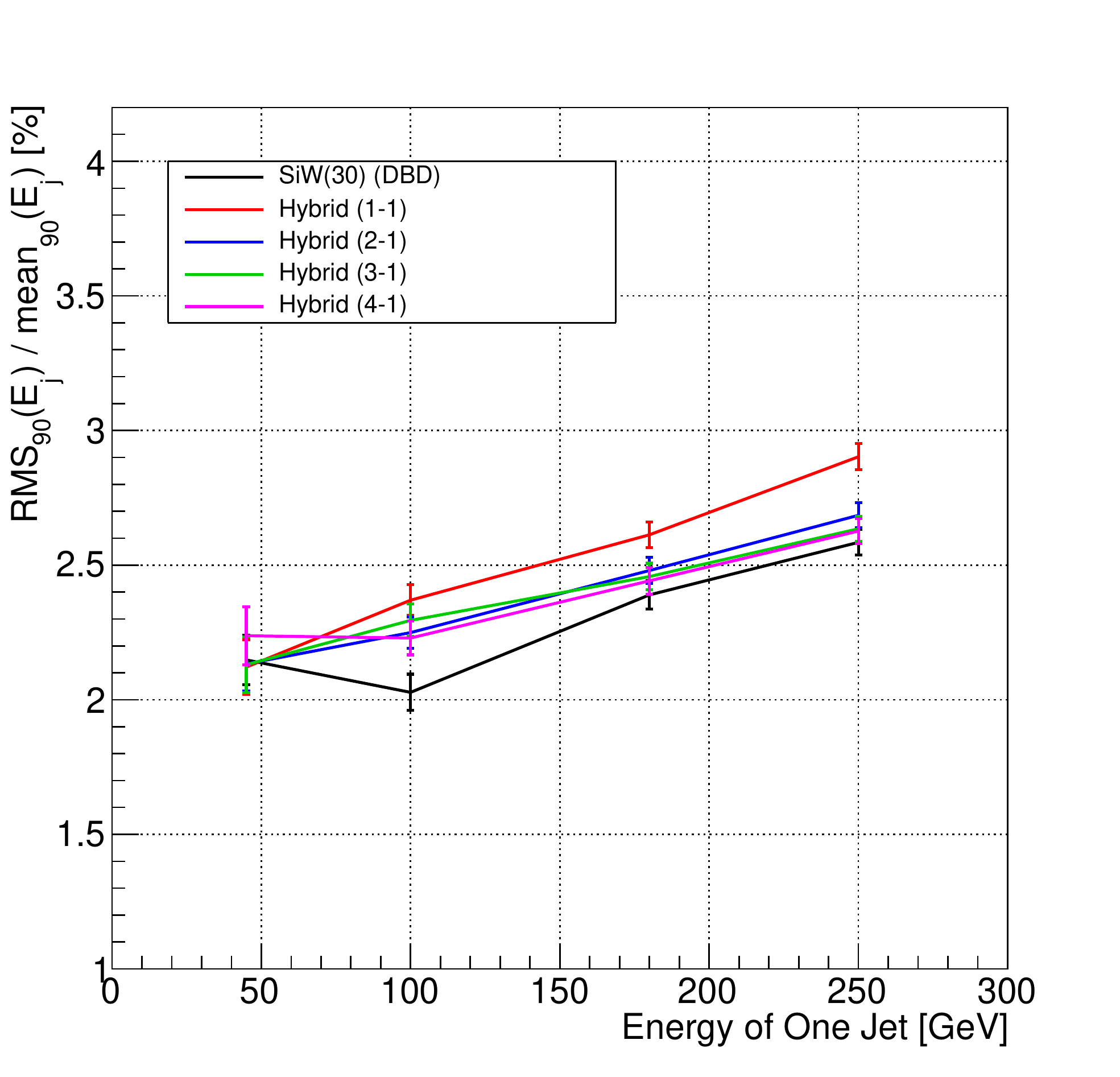}
\end{minipage}
\caption{Comparison of jet energy resolution (left) and confusion term (right) between 4 configurations and silicon ECAL in DBD}
\end{figure}

\section{Summary and Plan}

We studied the performance of various hybrid ECAL configurations for ILD using ILCSoft v01-16-02. The hybrid ECAL used silicon semiconductor in inner 14 layers and scintillator-tile in outer layers. We also added a requirement that whole thickness of absorber was fixed as $22.8X_0$. We changed the number of scintillator layers and a boundary between inner and outer region, then we got a result that an hybrid ECAL with 16 scintillator layers and a boundary of $11.2X_0$ had degradation of jet energy resolution by about 2.7\% compared with DBD silicon ECAL with 30\% cost reduction. Therefore, we guess that this hybrid ECAL is capable to resolve the cost issue of ILD in this study.

Our future plan is to use medium region with alternating structure of silicon and scintillator to partially reduce the effect of confusion. 

\section*{Acknowledgement}

The authors would like to thank the ILC Physics Working Group, ILD SiW-ECAL group and CALICE-Asia group for useful discussions. This work has been partially supported by JSPS Specially Promoted Research No. 23000002.

\end{document}